# Simulating is not always understanding

Lendert Gelens[1,*], Alejandro Fábregas-Tejeda[2], Grant Ramsey[2], Sylvia Wenmackers[2], and Bart Smeets[3]

[1]*Laboratory of Dynamics in Biological Systems, Department of Cellular and Molecular Medicine, KU Leuven, Herestraat 49, Leuven, 3000, Belgium*

[2]*Centre for Logic and Philosophy of Science, Institute of Philosophy, KU Leuven, Kardinaal Mercierplein 2, Leuven, Belgium*

[3]*MeBioS, Department of Biosystems, KU Leuven, Kasteelpark Arenberg 30, 3001, Leuven, Belgium*

**Corresponding author: lendert.gelens@kuleuven.be (L. Gelens)*

## Abstract

Computational models range from minimal representations to whole-cell simulations, and increasing complexity changes what they describe, predict, and explain. While adding detail may improve descriptive and predictive capabilities, as recent whole-cell simulations illustrate, it does not necessarily increase mechanistic understanding. What matters is not component count, but whether model parameters are sufficiently constrained by experimental data, and whether we can explain why the model behaves as it does. Agent-based models of cytoskeletal dynamics reveal rich self-organization with parameter spaces sufficiently small to explore systematically. When free parameters outrun the data available to constrain them, models become harder to interpret and to critically test. We argue that extracting mechanistic understanding from complex simulations requires comparison with simpler representations, dynamical analysis, and explicit model hierarchies that reveal which details matter for the behavior of interest.

**Glossary:**

**Adequacy-for-purpose view of models**: Stance that contends that judgements of a model's adequacy depend on how well it fulfils its intended use and meets associated epistemic standards.

**Agent-based models**: Models that simulate rule-following, interacting individual agents to show how local behaviors can give rise to collective, general patterns.

**Architectural (modeling) assumptions**: Assumptions about a model's components, structure, and posited relations among components.

**Bifurcation analysis**: A widespread method for identifying shifts in a system's dynamics as parameters vary (e.g., transitions between stable states and oscillatory behaviors).

**Bistable switch**: Representation of a system that can have two stable states under the same environmental conditions.

**Coarse-grained models**: Models that represent a target system at a larger scale than a more comprehensive, variable-rich representation.

**Continuum active matter descriptions**: Models that represent active matter systems (composed of intrinsically-driven, energy-consuming individual particles) using continuous fields.

**Control parameters**: Parameters whose variation can decisively alter a system's behavior and serve to determine transitions between dynamical regimes.

**Counterfactual scenarios**: Theoretical scenarios that allow to pose questions about what would have happened if things had been different (e.g., regarding a system's inputs, its key parameters, or typical background conditions).

**Free parameters**: Parameters whose (non-fixed) values must be estimated from empirical data.

**Functional-response models**: Models that track how a given variable or a selected process reacts to different input values.

**Mass-action models**: Models in which reaction rates depend on the concentrations of interacting components and kinetic constants.

**Mechanistic models**: Models that represent and foreground components of a system and the interactions through which they causally produce salient phenomena.

**Minimal sufficient representation**: The simplest representation of a model that retains the relevant features needed to answer a particular research question or elicit a particular focal behavior.

**Model hierarchies**: Sets of related models that differ in parameter complexity or detail. Comparing them (e.g., through model reduction) helps detect which features are necessary or sufficient to produce a phenomenon and which parameters the modeler can dispense with.

**Parameter non-identifiability**: A condition in which different combinations of parameters yield identical observables or indistinguishable predictions and thus a model's parameters cannot be uniquely and unambiguously established.

**Parameter sloppiness**: A property of complex models whereby some unconstrained parameter combinations strongly impact (collective) model outputs while many other parameters have little to no effect.

**Phase plane methods**: Methods that analyze differential equations by plotting their dynamics in a phase plane to identify fixed points, chart its stability, and uncover transitions between dynamical states.

**Sensitivity analysis**: Method for assessing how changes in parameters, inputs, or initial conditions sway model outputs and how well predictive robustness is maintained.

**Sloppy model**: A model whose predictions are robust yet highly sensitive to some parameter combinations and relatively insensitive to many others (whose values can dramatically vary by several orders of magnitude).

**Slow manifolds**: Regions of state space, derived through dimensionality reduction techniques, along which system dynamics develop slowly, facilitating its scientific study and overall tractability.

**Stiff parameter combinations**: Parameter combinations to which model outputs and predictions are highly sensitive to.

**Universality class**: A category of systems, tapped by minimal models, sharing the same large-scale behavior despite differences in microscopic details and instantiation.

**Vertex and foam models**: Models that construe cells as polygons or polyhedra sharing boundaries, commonly marshaled to probe how cell geometry and mechanical forces shape tissue organization.

**Whole-cell simulations**: Prediction-friendly and parameter-heavy computational models that are able to track individual molecular species through complete cell cycles, tightly coupling processes such as metabolism, gene expression, chromosome dynamics, and membrane growth.

## What is a model for?

The landmark 2012 **whole-cell simulation** (**see Glossary**) of *Mycoplasma genitalium* [14] and the 2026 spatially-resolved model of JCVI-syn3A [15], expanding the results from [41], are substantial technical achievements in contemporary cell biology. These simulations track individual molecular species through complete cell cycles, explicitly coupling metabolism, gene expression, chromosome dynamics, and membrane growth. Building such models forces explicit quantitative accounting of cellular processes and properties, identifies measurement gaps, and can expose inconsistencies between data sources and across biological domains [42, 43]. Likewise, whole-cell models can help discover properties that emerge where dynamically interacting processes intersect [44] and enable systematic, hard-to-run in the laboratory *in silico* genetic and environmental perturbations.

However, successfully simulating a system does not always increase our understanding of it. For instance, a simulation that reproduces an experimental observation confirms that the model is consistent with that observation, but does not establish that it captures the mechanism responsible, since different mechanisms can produce the same output. Models can describe what a system does, predict how it will behave under different conditions, or explain why it behaves as it does. These achievements are related, but not equivalent (Box 1). Mechanistic understanding requires identifying how components and interactions give rise to a behavior, securing generalizability under **counterfactual scenarios** (e.g., predicting what will happen when those components are changed through experimental manipulation), and explaining why the system behaves one way rather than another [8–11]. Models can appear explanatory without uniquely recovering the underlying causal mechanisms, for example because different parameter combinations can produce the same result (**parameter non-identifiability**) or because **architectural assumptions** built into the model generate behaviors that are absent from the biological phenomena being simulated.

A model — always a work in progress and open to revision — can contribute additional scientific insight when it does more than reproduce its inputs. It might predict an observation that was not used to construct it, reveal that a behavior requires a previously unsuspected interaction, or fail in a specific way when a component is removed, suggesting a causal role for that component. Model simplicity can be advantageous because a lean model may show that complex behavior can be reproduced from a much simpler system, *reducing* rather than adding complexity [12, 13]. These considerations are particularly timely as computational cell biology moves toward increasingly complex models, especially whole-cell simulations [6, 7].

Although issues such as the limitations of parameter inference [16–19], the value of minimal models [20–22], and the distinction between prediction and explanation [23, 24] have been separately discussed in the literature (e.g., in systems biology or in the philosophy of science), their implications for whole-cell modeling (and for cell biology more broadly) deserve renewed scrutiny. We provide an integrative theoretical and philosophical appraisal that brings different insights into dialogue to examine whether and how complex models and simulations in cell biology yield scientific understanding of living phenomena. We argue that increasing the completeness of such models does not necessarily increase mechanistic understanding, and that the field should devote as much attention to dissecting and explaining their behavior as to constructing ever more comprehensive simulations. As another contribution, we provide a few actionable suggestions on how this perspective could impact biological modeling practices and their interpretation: in particular, we call for building explicit model hierarchies through model reduction and complementing this with dynamical analysis.

**Box 1. Some uses of models and the adequacy-for-purpose view**

We can learn many things about particular target systems by constructing and manipulating models [1] that stand in for them (e.g., when direct access is curtailed)—a paradigmatic instance of what is called ‘surrogative reasoning.’ Models serve manifold epistemic roles in scientific practice [2], many of which, but not all of them, are tied with an underlying representationalist logic (viz., a model is assumed to represent its target system). First, a model can play a *descriptive* role, telling us what its target system (or selected aspects of it) is like. It can, for instance, systematize observations, display relationships among entities, and identify salient variables in a cellular process of interest. Second, a model can *explain* phenomena in diverse ways [3]. In cell biology, it is commonplace that explanatory models are able to show how entities, activities, and their organization produce a given outcome [4; but see 5]. A mechanistic model of a cellular process, for example, may specify how certain molecular components interact, how those interactions unfold over time or how specific they are, and how their organization generates a particular focal behavior of the target system. Third, models can aid in *prediction* [6], for instance, they can be useful to derive consequences from assumptions about a system. These consequences may concern future behavior, unobserved states, or counterfactual states (i.e., asking about what would have happened if things had been different, for instance, if a key variable were to be changed). Moreover, when competing hypotheses are formalized as alternative models, their predictions can be assessed against evidence and thus help to discern which assumptions are consequential. Fourth, models can also be employed with *exploratory* functions [7], without appealing to hypothesis testing, mechanistic grounding, or connection to explanation overall.

The adequacy-for-purpose view of models [46] that we advocate here dictates that the criteria for determining a model’s success must be attuned depending on which function it is performing: description, explanation, prediction, or exploration. For instance, if the goal of a model is forecasting, different epistemic standards apply than when it serves descriptive or causal-mechanistic purposes. The adequacy-for-purpose view invites us to reflect on why we build particular models (i.e., what we intend to achieve) and on the limitations they entail *vis-à-vis* other modeling purposes. However, not all modeling purposes contribute equally to understanding: explanation, foremost among them, is particularly important as an epistemic achievement in a science such as cell biology.

**Box 1. Some uses of models and the adequacy-for-purpose view**

## Identifiability and interpretability

The appeal of detailed **mechanistic models** is clear: *prima facie*, if a model includes more of the known biology, its behavior should be more realistic and its predictions more reliable. However, although model complexity can improve predictive power, it may undermine scientific understanding. What determines whether a model generates insight is not simply how many components it contains. Instead, two important factors are whether (1) its **free parameters** can be constrained by the available experimental data (*identifiability* [25–28]) and (2) the relationship between those parameters and the resulting behavior can be understood (*interpretability* [29–31]). These factors are distinct: an identifiable model can still be a black box, and even a **sloppy model** with many poorly constrained parameters can be interpretable if the stiff parameter combinations that determine its behavior can be identified. Sloppiness describes unequal sensitivities to parameter combinations and does not necessarily imply non-identifiability [27, 32, 33]. A model that fails on both counts offers little mechanistic understanding (Figure 1). The tension between adding realism and preserving these properties echoes Levins' classic argument that generality, realism, and precision in modeling come with trade-offs, so they cannot all be maximized at once [34], as well as Weisberg's account of different representational ideals in modeling [12].

This distinction is illustrated by a class of models that are large in spatial scale and component number but use relatively few parameters. **Agent-based** simulations of cytoskeletal dynamics, in which hundreds of thousands of filaments and motor proteins interact according to a small number of physically grounded rules, have revealed how contractile rings, mitotic spindles, and cortical flows self-organize through local interactions [35]. These models can be amenable to systematic analysis because relatively few governing parameters, often independently measurable in single-molecular experiments, control their collective behavior.

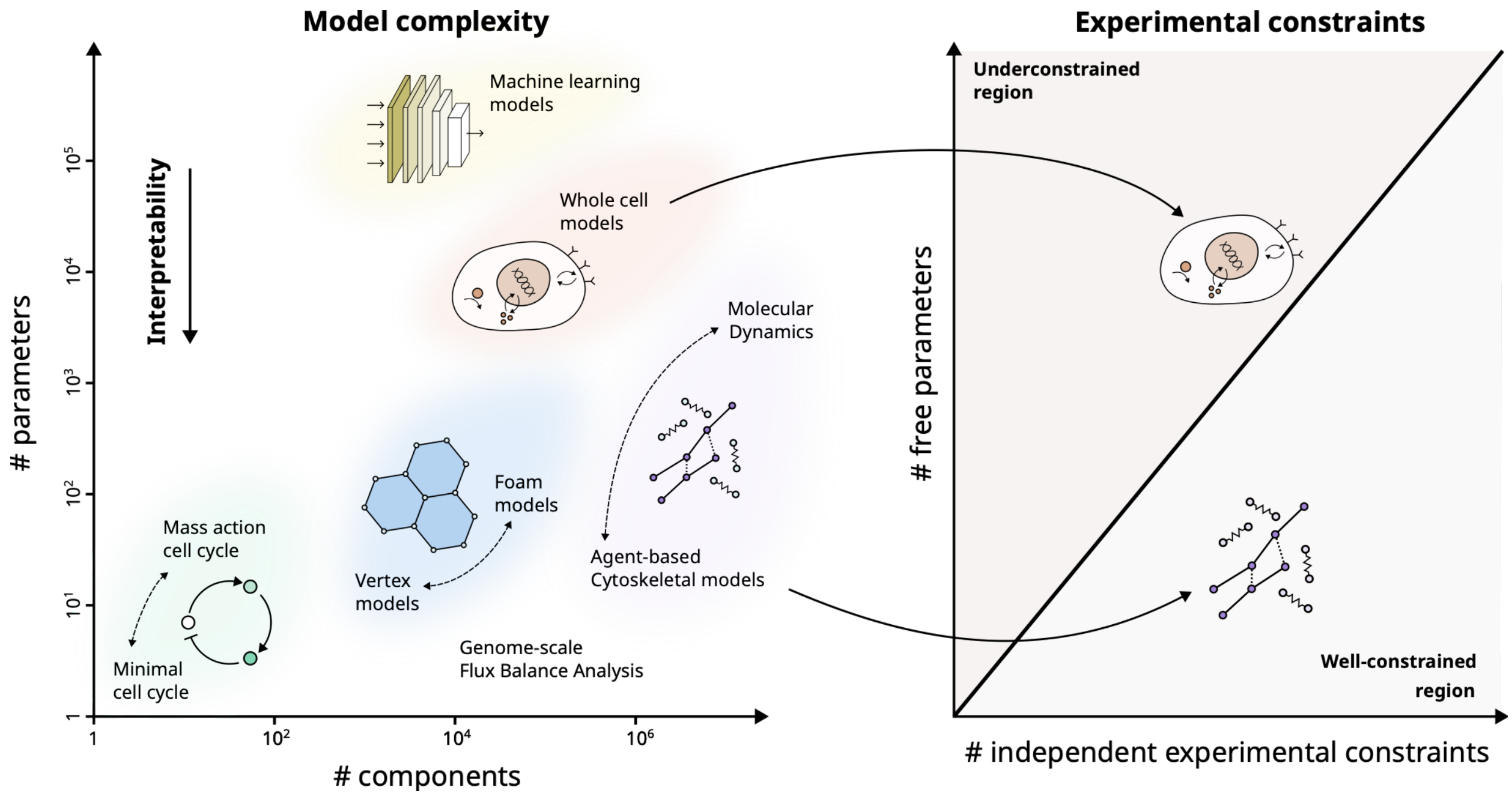

Figure 1: **Schematic representation of model types in terms of model complexity and experimental constraints.** These conceptual maps are not based on precise measurements. On the left, model types are represented according to component and parameter count; increasing parameter number can reduce interpretability. On the right, two of the same model types are shown according to how well their free parameters are constrained by independent experimental measurements, distinguishing well-constrained from underconstrained models. This schematic distinction is not a formal criterion for identifiability, which also depends on model structure and the information contained in the measurements. A model can contain many components yet remain well constrained if it is governed by a small number of physically grounded rules. In particular, agent-based cytoskeletal models can remain well-constrained despite differing widely in component count, whereas the free parameters of whole-cell models can substantially outnumber the available experimental constraints.

**Vertex and foam models** of epithelial tissues achieve the same epistemic economy: three or four parameters governing cell area elasticity, perimeter tension, and adhesion energy can reproduce tissue fluidity, jamming transitions, and the mechanics of cell intercalation during morphogenesis [36]. **Continuum active matter descriptions** of cell monolayers, with similarly few parameters, predict large-scale flows, topological defects in cell orientation, and the statistics of collective migration [37]. These models are not simple in the sense of being small; rather the mapping from parameters to behavior can remain sufficiently low-dimensional and tractable to investigate systematically how particular interactions generate emergent behaviors (for discussion, see also [38]; for a contrasting viewpoint, see [39]).

Current whole-cell models [e.g., 14, 15], even if predictively powerful [40] and exhibiting many advantages, present a sharp contrast. The parameter space of these models is too large to be explored exhaustively. Many parameters rely on estimates transferred from related model organisms or different experimental conditions [45], and comparatively little work has been done to understand why the simulations behave as they do. This distinction can be framed in terms of **adequacy-for-purpose** [46; Box 1]: these models may be adequate for descriptive purposes (e.g., integrating data), exposing measurement gaps, and generating hypotheses, but these achievements do not by themselves establish mechanistic explanations [11]. Thus, although creating such models requires considerable knowledge of cellular processes, it is not clear that the simulation *itself* generates novel understanding. We stress that our claim is contingent, not principled. We do not argue that whole-cell models are *inherently* incapable of yielding understanding, but that they have not *yet* been subjected to the systematic analysis needed to establish how their simulated behaviors arise.

## When parameters outrun constraints

When building understanding-conducing models, the difficulty is not scale per se: what matters is not the dimensionality of the state space, but the *dimensionality of the parameter space* that governs it. Problems arise when free parameters proliferate faster than the experimental constraints available to determine them, and when no systematic effort is made to map the relationship between parameters and the system's behavior [47]. As the number of unconstrained free parameters grows, they increasingly occur in combinations that cannot be disentangled using the available measurements. The parameter space may develop large, flat regions where many different parameter combinations fit the data equally well [25, 48]. A related but distinct property is **parameter sloppiness**, which is widespread in systems biology models: when parameters are varied systematically, a small number of combinations strongly affect predictions, whereas most others can vary by orders of magnitude with little detectable effect on observable outputs [25, 26, 45].

In the systems biology literature, parameter sloppiness is frequently presented as good news: it is precisely *because* predictions depend on only a few **stiff parameter combinations** that models can forecast reliably without requiring measurements of most individual rates [25]. This reading is correct on its own terms, but it concerns prediction, not causation, and this distinction is central to our argument. The property that makes a sloppy model robustly predictive can also weaken the link between a good fit and the underlying mechanism [49]. If observable behavior is insensitive to changes in most parameter combinations, reproducing that behavior tells us little about the values of those parameters and, therefore, does not establish whether the mechanisms they encode operate in the cell. Such insensitivity may also reflect biological degeneracy, whereby different molecular configurations generate similar functional behavior. In these cases, understanding may require identifying common dynamical principles rather than a unique microscopic mechanism [50–52]. Sloppiness is thus a friend of prediction but can also pose a challenge to mechanistic understanding [11]. A model can be trustworthy as a forecasting tool and uninformative as a causal explanation; conflating these two functions is the error we are warning against.

Box 2 illustrates how comparing models at different levels of detail can help address this challenge, using a biological oscillation: the cell cycle [53, 54]. A detailed **mass-action model** and a minimal **functional-response model** generate essentially the same oscillations, but represent the regulatory mechanism at different levels. The detailed model describes the molecular interactions underlying a **bistable switch**, whereas the minimal model directly incorporates its experimentally measured functional response. Comparing these representations helps reveal which aspects of the regulatory mechanism are needed to account for cell cycle oscillations, and which require a more detailed molecular description. When a complex model reproduces an observation, this establishes consistency between the model and the data, which is necessary but not sufficient for the model to be correct and insightful. It does not establish that the model has identified the mechanism responsible for the phenomenon, because other models with different mechanisms may be equally consistent with the data. Mechanistic claims from models with many unconstrained parameters therefore require additional support, including systematic comparison with simpler representations of the same system. The next step is to identify which components, parameters, and transitions actually control the behavior.

## Box 2. Cell cycle models illustrate the value of model hierarchies

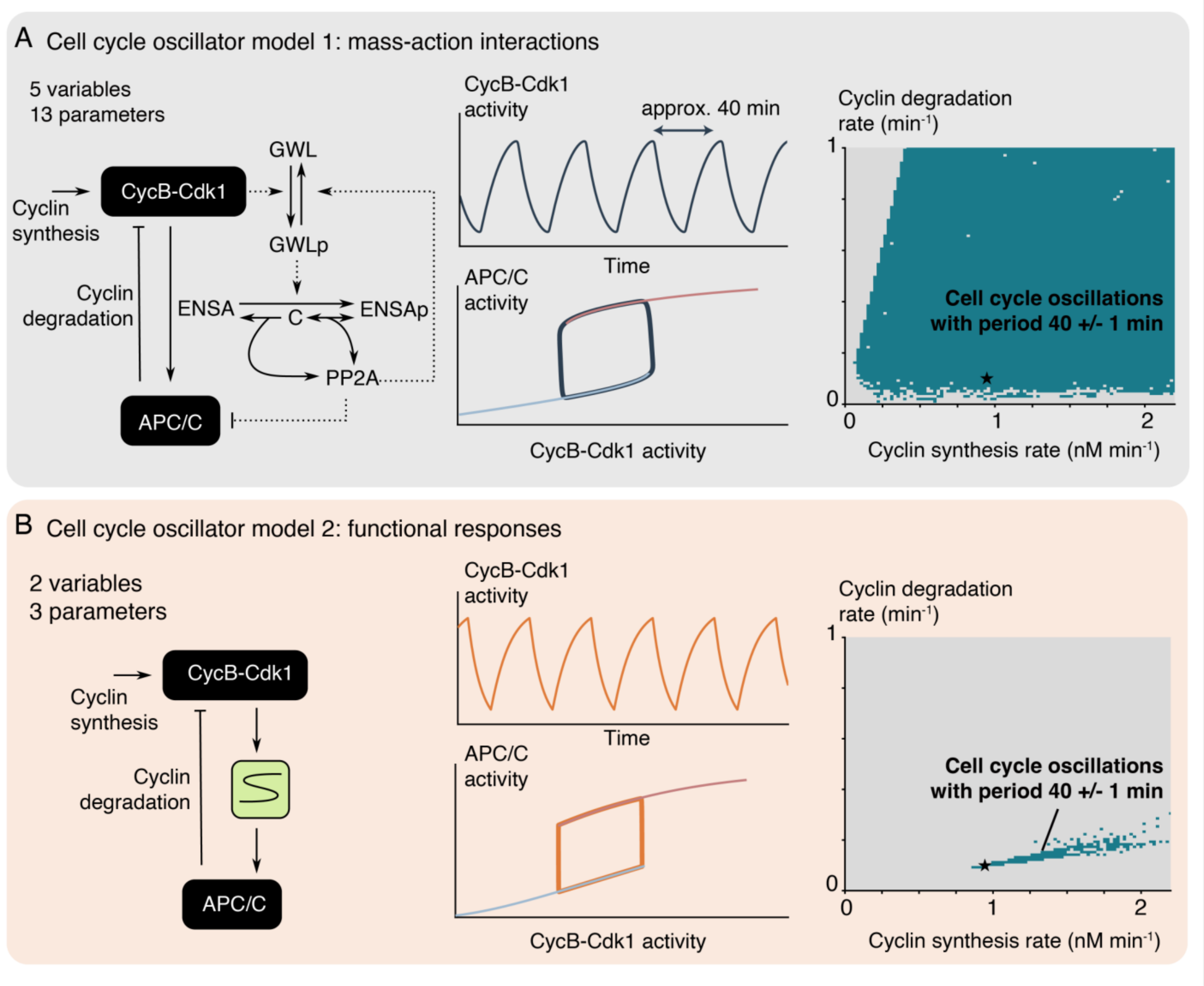

***Figure I. Detailed and reduced models of the cell cycle oscillator.*** *(A) Detailed mass-action model with five dynamical variables and 13 free parameters, explicitly representing the Cyclin B–Cdk1, Greatwall kinase (GWL), ENSA, and PP2A regulatory circuit. (B) Reduced functional-response model with two dynamical variables and three free parameters, replacing the GWL–ENSA–PP2A subnetwork with an experimentally measured bistable response. Both models reproduce similar oscillations (middle, top) and phase-plane dynamics (middle, bottom), including the transition between interphase and mitosis. The right panels show combinations of Cyclin B synthesis and degradation rates compatible with an oscillation period of 40 ± 1 min (colored regions), based on 5000 randomly sampled internal parameter sets in (A) and the independently constrained functional response in (B). Stars indicate the parameter values used for the illustrated trajectories. The broader compatible region in (A) reflects the possibility of compensating changes in internal kinetic parameters, whereas the measured functional response restricts this freedom in (B). Models are based on [53].*

Multiple models of varying complexity have been developed for the embryonic cell cycle of *Xenopus laevis*, illustrating how different levels of description can reveal complementary aspects of the same regulatory mechanism. Periodic Cyclin B-Cdk1 (CycB-Cdk1) activity drives entry into mitosis, whereas Cyclin B degradation by the anaphase-promoting complex/cyclosome (APC/C), drives the cell back to interphase. A bistable feedback module involving Greatwall kinase (GWL), ENSA, and PP2A makes this transition switch-like. Coupling this switch to Cyc B synthesis and degradation produces sustained CycB-Cdk1 oscillations [53, 54].

The detailed model in Figure I(A) represents this circuit using mass-action kinetics, with five dynamical variables and 13 free parameters. The minimal model in Figure I(B) replaces the PP2A–ENSA–GWL subnetwork with the experimentally measured S-shaped relationship between CycB-Cdk1 and APC/C activity. Because this functional response is fixed independently by bistability measurements, only three free parameters remain. Both models reproduce oscillations with similar periods and phase-plane structures.

**Box 2. Cell cycle models illustrate the value of model hierarchies**

The two models address different levels of mechanistic explanation. The detailed model represents the molecular interactions underlying the bistable switch, whereas the minimal model captures their experimentally measured collective response. At the level of cell cycle dynamics, this reduction clarifies how bistability, together with Cyclin B synthesis and degradation, generates oscillations. Explaining the molecular origin of bistability, however, requires the more detailed representation.

The models also differ in how observations constrain their parameters. In the detailed model, changes in internal kinetic rates can compensate for changes in Cyclin B synthesis and degradation, whereas these rates are more tightly constrained in the reduced model because the bistable response is fixed independently by experiment. This does not imply that simpler models are intrinsically more identifiable: additional measurements could also constrain the detailed model. Rather, the comparison illustrates how matching the level of description to the available empirical information can facilitate mechanistic analysis.

Together, these representations form a hierarchy connecting molecular interactions, an effective regulatory response, and cell cycle oscillations. Comparing them helps determine which level of detail is needed to explain a particular behavior, without implying that omitted molecular interactions are biologically irrelevant.

## What it takes to gain understanding through modeling

With regard to simulations, it is important to distinguish between the *sense* of understanding — a subjective, cognitive state in which a model *feels* adequate or appears to point in the right direction — and understanding proper, an epistemic state that allows an individual or scientific community to make phenomena intelligible and grasp how and why they occur [55]. (For a discussion of understanding proper, see [56, 57].) A smoothly running simulation can therefore create an illusion of understanding when its apparent realism exceeds what can actually be explained [58]. We argue that two practices, often absent from the analysis of complex models in cell biology, help distinguish a model that adds to our understanding from one that merely reproduces observations. Neither is sufficient on its own, and a model that provides both is considerably more likely to support causal conclusions than one that offers neither — but the criteria are better understood as a standard of rigor than as a checklist, since partial progress on either dimension is still worthwhile scientific progress.

The first practice is building explicit **model hierarchies** that use both systematic reduction and increasing levels of detail to investigate which components and interactions contribute to a given behavior. Models can be built from minimal representations upward or simplified from more comprehensive descriptions, with experiments guiding the appropriate level of detail. At each step, two questions should be asked: does the simpler model reproduce the target behavior, and, if so, what does the added complexity contribute? If the reduced model succeeds, the omitted components are not required to reproduce that behavior within the reduced description, although they may still contribute causally in the biological system. If it fails, the failure identifies interactions necessary within the model, providing hypotheses that can be tested experimentally. Such comparisons also reveal when multiple microscopic representations of cellular states and processes produce the same effective dynamical description —in other words, when the details of implementation are not needed to account for a particular behavior and a coarser representation captures the essential physics. This is how a minimal model can have real explanatory power. On the account of Batterman and Rice, such models explain not by accurately mirroring their target but by belonging to the same **universality class**: a family of microscopically distinct systems can display the same large-scale behavior because their distinguishing details are irrelevant to that behavior [59], however painstakingly characterized [60]. Unlike simplifications made merely for computational convenience, such reductions can reveal why certain microscopic details do not affect the behavior of interest [4, 59]. The corollary for cell biology is that piling in biochemical detail can obscure causal explanations when that detail does not influence the behavior of interest. This point connects to the broader argument that greater mechanistic completeness does not automatically provide greater explanatory power [61].

In many biological cascades, removing intermediate steps changes the microscopic implementation but preserves the effective input–output response. The explanatory task is then not to retain every biochemical detail, but to identify the **minimal sufficient representation** [62]. Without this iterative decomposition, a complex model that simulates realistic behavior provides no guarantee that it does so for the right reasons. The cell cycle models in Box 2 illustrate how comparing molecular and effective descriptions can clarify what each level contributes to explaining the same behavior.

The agent-based cytoskeletal literature offers a template. The transition from a disordered cytoskeletal network to an aster or contractile ring can be traced through a hierarchy from single-filament mechanics, through pairwise motor-filament interactions, to collective behavior. Each level can identify which details matter, which can be **coarse-grained**, and which effective interactions govern the next scale [35]. For example, one can ask whether self-organized asters require active motor crosslinking or whether passive crosslinkers suffice. The model can give a crisp answer because its parameter space is small enough to explore systematically. The same analytical move is much harder in a whole-cell model where thousands of parameters may change simultaneously as components are added or removed.

Comprehensive models can nevertheless help identify which simplifications are justified, just as minimal models can be extended when observations demand greater detail. Our claim is not that comprehensive models are dispensable, but that their construction cannot be the terminus. A whole-cell model that reproduces cellular behavior, but is never reduced or subjected to analysis of which parameters and interactions actually control that behavior, has supplied the raw material for understanding without yet delivering the understanding itself.

The second practice is identifying **control parameters**, bifurcations, and the boundaries between robust and fragile regimes. The two practices are mutually reinforcing: dynamical analysis is often precisely how one discovers what can safely be discarded during model reduction. For example, sensitivity or bifurcation analysis can reveal which parameters and interactions control a behavior and which can be collapsed without losing that behavior. Model reduction and dynamical analysis are thus two faces of the same investigative move rather than two separate tasks. Asking whether a model reproduces an observation is a static question. Asking why it produces that behavior, what makes the behavior robust or fragile, where the bifurcations lie, and what controls transitions between different behavioral regimes is a dynamical question — and it is this type of analysis that generates scientific understanding. The minimal modeling tradition has developed extensive tools for this purpose, including **bifurcation analysis**, **sensitivity analysis**, **phase plane methods**, and reduction techniques that identify **slow manifolds** and control parameters [29, 31]. These methods are more difficult to apply systematically to large-scale models, partly because the computational cost of each simulation makes extensive parameter-space exploration prohibitive, and partly because the culture around such models has focused more on construction and reproduction rather than analysis. Such analysis can move beyond reproducing observed dynamics to identifying the conditions under which different behaviors arise. The vertex model literature offers an example at the tissue scale: mapping the jamming transition as a function of the ratio of cell perimeter to cell area turned a simulation into a predictive phase diagram for tissue mechanics [63]. Here, the move from reproduction to analysis generates experimentally testable predictions. Performing comparable analyses on whole-cell models is much more challenging: thousands of uncertain parameters, strong coupling between cellular processes, and computationally expensive simulations make exhaustive exploration impractical. Nevertheless, targeted sensitivity analysis and reduction of specific modules can provide starting points for investigating which interactions control a particular behavior.

None of this requires abandoning large-scale models. On the contrary, it requires employing them differently: as laboratories for decomposition rather than demonstrations of comprehensiveness (see also [60]). Research centers have made substantial investments in simulation infrastructure; realizing its scientific value requires complementary investment in the methods needed to extract understanding from these models.

## Concluding remarks

Here, we have recounted cases where model complexity can obscure — rather than reveal — biological phenomena. As we argued, what matters is not how many components a model contains, but whether its free parameters are constrained by available data and whether the relationship between those parameters and the resulting behavior can be understood. Agent-based models of cytoskeletal self-organization and vertex models of tissue mechanics demonstrate that large-scale spatial simulations can generate deep insight when built on parameter-lean foundations and analyzed systematically. Whole-cell models, despite their impressive scope and technical prowess, have not yet achieved the same standard, in part because systematic approaches to extracting mechanistic understanding from them remain limited. We have offered some suggestions for how this could be achieved (see **Outstanding Questions**).

Changing the modeling *ethos* requires treating large-scale simulations not as endpoints but as starting points of biological investigations: structured representations of current knowledge that can yield mechanistic understanding when compared with simpler models, interrogated dynamically, and organized into explicit model hierarchies that reveal which details matter for particular behaviors. Understanding can also emerge from grasping how different levels of description relate to one another, rather than from any single model in isolation [59, 11]. This is an argument for more, rather than less, ambition in biological modeling, and for directing that ambition toward the moves that convert simulation into understanding-conducing explanation [31].

The arrival of machine learning approaches that predict cellular behavior at scale and integrate a wide array of biological data types [64, 65] makes the analytical demand on mechanistic (and theoretically insightful) models even more pressing [66]. Prediction is increasingly available without mechanistic understanding. The challenge is not simply to simulate increasingly complete cells, but to understand why these simulations behave as they do, which mechanisms their predictions depend on, and which details can be discarded without losing explanatory power. This, we believe, will push the frontiers of cell biology even further.

### Data availability statement

The numerical codes to reproduce Figure I in Box 2 in this work are openly available in GitLab [67].

### Acknowledgments

All authors acknowledge funding by the KU Leuven Research Fund (grant number IDN/25/007).

### Competing interests statement

The authors declare no competing interests.

### Declaration of generative AI and AI-assisted technologies in the manuscript preparation process

During the preparation of this work, the author(s) used Claude to check grammar and writing. The author(s) reviewed and edited the output as needed and take full responsibility for the content of the published article.

## References

[1] Margaret Morrison and Mary S Morgan. Models as mediating instruments. *Ideas in Context*, 52:10–37, 1999.

[2] Stephen M Downes. Scientific models. *Philosophy Compass*, 6(11):757–764, 2011.

[3] Alisa Bokulich. How scientific models can explain. *Synthese*, 180(1):33–45, 2011.

[4] William Bechtel. Mechanism and biological explanation. *Philosophy of science*, 78(4):533–557, 2011.

[5] Derek John Skillings. Mechanistic explanation of biological processes. *Philosophy of Science*, 82(5): 1139–1151, 2015.

[6] Alkistis Elliott-Graves. What are general models about? *European Journal for Philosophy of Science*, 12(4):74, 2022.

[7] Michela Massimi. Two kinds of exploratory models. *Philosophy of Science*, 86(5):869–881, 2019.

[8] William Bechtel and Adele Abrahamsen. Dynamic mechanistic explanation: computational modeling of circadian rhythms as an exemplar for cognitive science. *Studies in History and Philosophy of Science*, 41:321–333, 2010.

[9] Carl F. Craver. When mechanistic models explain. *Synthese*, 153:355–376, 2006.

[10] Judea Pearl. *Causality: Models, Reasoning and Inference*. Cambridge University Press, Cambridge, 2nd edition, 2009.

[11] Yasha Rohwer and Collin Rice. How are models and explanations related? *Erkenntnis*, 81(5):1127–1148, 2016.

[12] Michael Weisberg. Three kinds of idealization. *Journal of Philosophy*, 104:639–659, 2007.

[13] Tarja Knuuttila. Modelling and representing: an artefactual approach to model-based representation. *Studies in History and Philosophy of Science*, 42:262–271, 2011.

[14] Jonathan R. Karr, Jayodita C. Sanghvi, Derek N. Macklin, Miriam V. Gutschow, Jared M. Jacobs, Benjamin Bolival, Nacyra Assad-Garcia, John I. Glass, and Markus W. Covert. A whole-cell computational model predicts phenotype from genotype. *Cell*, 150:389–401, 2012.

[15] Zane R. Thornburg, Ansel Maytin, Jonghan Kwon, Troy A. Brier, Benjamin R. Gilbert, Enguang Fu, Yuqing L. Gao, James Quenneville, Tejal Wu, Handuo Li, et al. Bringing the genetically minimal cell to life on a computer in 4d. *Cell*, 189:2582–2597, 2026.

[16] Andreas Raue, Clemens Kreutz, Thomas Maiwald, Julie Bachmann, Marcel Schilling, Ursula Kling-Müller, and Jens Timmer. Structural and practical identifiability analysis of partially observed dynamical models by exploiting the profile likelihood. Bioinformatics, 25(15):1923–1929, 2009.

[17] Oana-Teodora Chis, Julio R Banga, and Eva Balsa-Canto. Structural identifiability of systems biology models: a critical comparison of methods. PloS one, 6(11):e27755, 2011.

[18] Preston, S.P. *et al.* (2025) Think before you fit: Parameter identifiability, sensitivity and uncertainty in systems biology models. *Current Opinion in Systems Biology* 42, 100563

[19] Gasior, K.I. (2026) Determining parameter impact in systems biology models via sensitivity analysis: a comparative approach. *npj Syst Biol Appl* 12, 143

[20] Robert W Batterman. Asymptotics and the role of minimal models. *The British Journal for the Philosophy of Science*, 53(1):21–38, 2002.

[21] Till Grüne-Yanoff. Learning from minimal economic models. Erkenntnis, 70(1):81–99, 2009.

[22] Povich, M. (2018) Minimal Models and the Generalized Ontic Conception of Scientific Explanation. *The British Journal for the Philosophy of Science* 69, 117–137

[23] Eric Winsberg. Science in the age of computer simulation. University of Chicago Press, 2010.

[24] Weber, E. *et al.* (2013) *Scientific Explanation*, Springer Netherlands.

[25] Ryan N. Gutenkunst, Joshua J. Waterfall, Fergal P. Casey, Kevin S. Brown, Christopher R. Myers, and James P. Sethna. Universally sloppy parameter sensitivities in systems biology models. *PLoS Computational Biology*, 3:e189, 2007.

[26] Kamil Erguler and Michael P. H. Stumpf. Practical limits for reverse engineering of dynamical systems: a statistical analysis of sensitivity and parameter inferability in systems biology models. *Molecular BioSystems*, 7(5):1593–1602, 2011.

[27] Oana-Teodora Chis, Alejandro F Villaverde, Julio R Banga, and Eva Balsa-Canto. On the relationship between sloppiness and identifiability. *Mathematical biosciences*, 282:147–161, 2016.

[28] Javier Aguilar, Miguel A Muñoz, and Sandro Azaele. Limits of inference in complex systems: When stochastic models become indistinguishable. Physical Review X, 16(3):031015, 2026.

[29] John J. Tyson and Bela Novak. A dynamical paradigm for molecular cell biology. *Trends in Cell Biology*, 30:504–515, 2020.

[30] Emily Sullivan. Understanding from machine learning models. *British Journal for the Philosophy of Science*, 73(1):109–133, 2022.

[31] Prokop, B. and Gelens, L. (2026) Data-driven discovery of dynamical models in biology. *Nat Rev Phys* 8, 561–578

[32] Joshua F Apgar, David K Witmer, Forest M White, and Bruce Tidor. Sloppy models, parameter uncertainty, and the role of experimental design. Molecular BioSystems, 6(10):1890–1900, 2010.

[33] Christian Tönsing, Jens Timmer, and Clemens Kreutz. Cause and cure of sloppiness in ordinary differential equation models. Physical Review E, 90(2):023303, 2014.

[34] Richard Levins. The strategy of model building in population biology. *American Scientist*, 54(4): 421–431, 1966.

[35] François J. Nédélec, Thomas Surrey, Anthony C. Maggs, and Stanislas Leibler. Self-organization of microtubules and motors. *Nature*, 389:305–308, 1997.

[36] Silvanus Alt, Poulami Ganguly, and Guillaume Salbreux. Vertex models: from cell mechanics to tissue morphogenesis. *Philosophical Transactions of the Royal Society B: Biological Sciences*, 372(1720): 20150520, 2017.

[37] M. Cristina Marchetti, Jean-François Joanny, Sriram Ramaswamy, Tanniemola B. Liverpool, Jacques Prost, Madan Rao, and R. Aditi Simha. Hydrodynamics of soft active matter. *Reviews of Modern Physics*, 85:1143, 2013.

[38] Julie Schweer and Marcus Elstner. On the explanatory power of atomistic simulations. *European Journal for Philosophy of Science*, 15:45, 2025.

[39] Florian J. Boge. How to infer explanations from computer simulations. *Studies in History and Philosophy of Science Part A*, 82:25–33, 2020.[40] Lydia Freddolino and Saeed Tavazoie. Beyond homeostasis: a predictive-dynamic framework for understanding cellular behavior. *Annual review of cell and developmental biology*, 28(1):363–384, 2012.

[41] Thornburg, Z.R. *et al.* (2022) Fundamental behaviors emerge from simulations of a living minimal cell. *Cell* 185, 345-360.e28

[42] Wendy S. Parker. Evidence and knowledge from computer simulation. *Erkenntnis*, 87:1521–1538, 2022.

[43] Macklin, D.N. *et al.* (2020) Simultaneous cross-evaluation of heterogeneous *E. coli* datasets via mechanistic simulation. *Science* 369, eaav3751

[44] Fridolin Gross and Sara Green. The sum of the parts: Large-scale modeling in systems biology. Philosophy, Theory, and Practice in Biology, 9(10), 2017. doi: 10.3998/ptb.6959004.0009.010.

[45] Ann C. Babtie and Michael P. H. Stumpf. How to deal with parameters for whole-cell modelling. *Journal of The Royal Society Interface*, 14(133):20170237, 2017.

[46] Wendy S. Parker. Model evaluation: an adequacy-for-purpose view. *Philosophy of Science*, 87(3): 457–477, 2020.

[47] Lu Qiao et al. The evolution of systems biology and systems medicine: from mechanistic models to uncertainty quantification. *Annual Review of Biomedical Engineering*, 27:425–447, 2025.

[48] Julio R. Banga and Alejandro F. Villaverde. Mechanistic dynamic modelling of biological systems: the road ahead. *Current Opinion in Systems Biology*, 42:100553, 2025.

[49] Hans Hasse and Johannes Lenhard. Boon and bane: On the role of adjustable parameters in simulation models. In Mathematics as a tool: Tracing new roles of mathematics in the sciences, pages 93–115. Springer, 2017.

[50] Bryan C Daniels, Yan-Jiun Chen, James P Sethna, Ryan N Gutenkunst, and Christopher R Myers. Sloppiness, robustness, and evolvability in systems biology. Current opinion in biotechnology, 19(4):389–395, 2008.

[51] Andreas Wagner. Robustness and evolvability: a paradox resolved. Proceedings of the Royal Society B: Biological Sciences, 275(1630):91–100, 2008.

[52] Larissa Albantakis, Christophe Bernard, Naama Brenner, Eve Marder, and Rishikesh Narayanan. The brain's best kept secret is its degenerate structure. The Journal of neuroscience, 44(40):e1339242024, 2024.

[53] Jolan De Boeck, Jan Rombouts, and Lendert Gelens. A modular approach for modeling the cell cycle based on functional response curves. *PLoS Computational Biology*, 17:e1009008, 2021.

[54] Julia Kamenz, Lendert Gelens, and James E. Ferrell. Bistable, biphasic regulation of PP2A-B55 accounts for the dynamics of mitotic substrate phosphorylation. *Current Biology*, 31(4):794–808.e6, 2021.

[55] Jaakko Kuorikoski. Simulation and the sense of understanding. In Paul Humphreys and Cyrille Imbert, editors, *Models, Simulations, and Representations*, pages 168–187. Routledge, New York, 2013.

[56] Michael Strevens. No understanding without explanation. *Studies in History and Philosophy of Science*, 44:510–515, 2013.

[57] Henk W. de Regt. *Understanding Scientific Understanding*. Oxford University Press, New York, 2017.

[58] Petri K Ylikoski. The illusion of depth of understanding in science. *In Scientific understanding: Philosophical perspectives*, pages 100–119. University of Pittsburgh press, 2009.

[59] Robert W. Batterman and Collin C. Rice. Minimal model explanations. *Philosophy of Science*, 81(3): 349–376, 2014.

[60] Kunihiko Kaneko. Complex versus complicated systems biology, universality versus detailed modeling. *Current Opinion in Systems Biology*, 44:100589, 2026.

[61] Carl F. Craver and David M. Kaplan. Are more details better? on the norms of completeness for mechanistic explanations. *British Journal for the Philosophy of Science*, 71(1):287–319, 2020.

[62] Ingo Brigandt. Beyond reduction and pluralism: toward an epistemology of explanatory integration in biology. *Erkenntnis*, 73:295–311, 2010.

[63] Reza Farhadifar, Jens-Christian Roper, Benoit Aigouy, Suzanne Eaton, and Frank Jülicher. The influence of cell mechanics, cell–cell interactions, and proliferation on epithelial packing. *Current Biology*, 17: 2095–2104, 2007.

[64] Huasen Jiang, Xiaoyu Huang, Xiangpeng Bi, Wenjian Ma, Haibo Ni, Zhiqiang Wei, Pin Sun, Henggui Zhang, and Shugang Zhang. Artificial intelligence-enabled multi-scale virtual cell: perspective, challenges, and opportunities. *Briefings in Bioinformatics*, 27(bbag104), 2026.

[65] Haotian Cui, Alejandro Tejada-Lapuerta, Maria Brbić, Julio Saez-Rodriguez, Simona Cristea, Hani Goodarzi, Mohammad Lotfollahi, Fabian J. Theis, and Bo Wang. Towards multimodal foundation models in molecular cell biology. *Nature*, 640:623–633, 2025.

[66] Ball, Philip. “Should Biology Put Complexity First?” *Cell Systems* 16, no. 2 (2025).

[67] Lendert Gelens. Gitlab repository. https://gitlab.kuleuven.be/gelenslab/publications/2026_Gelens_perspective_simulating_and_understanding.git, 2026.